\documentclass[final,5p,times,twocolumn,authoryear]{elsarticle}

\usepackage{amssymb}
\usepackage{amsmath}
\usepackage{lipsum}
\usepackage{mathrsfs}

\journal{International Journal of Geometric Methods in Modern Physics}

\begin{document}

\begin{frontmatter}




\title{Degenerate wormhole as a topological oscillator}

%% \\ I. Radial dynamics

\author{Juri Dimaschko}
\address{Technische Hochschule Lübeck, Mönkhofer Weg 239, 23562 Lübeck, Germany}

\begin{abstract}
A spherically symmetric, matter-free degenerate wormhole is considered as a gravitational system with a single collective degree of freedom — the radius of its throat. The classical dynamics is determined by applying the equivalence principle to the motion of the throat, while canonical quantization reduces the dynamics to a spectral problem with the dynamical symmetry group \(SU(1,1)\).  The exact solution of this spectral problem reveals two qualitatively different regimes. For sub-Planckian masses, the quantum system exhibits Bohr-type states characterized by a gravitational Bohr radius \(\hbar^2/(GM^3)\). For super-Planckian masses, the quantum system exhibits a semiclassical regime represented by the Einstein-Rosen wormhole with Schwarzschild gravitational radius \(2GM/c^2\). Thus, the model connects the familiar quantum bound-state behavior with the semiclassical geometry of a gravitational horizon.

\end{abstract}




\end{frontmatter}

\tableofcontents

\section{Introduction}

Astrophysical black holes are well established as compact objects whose external spacetime is accurately described by Schwarzschild or Kerr vacuum geometry. However, the existence of an event horizon has not been directly confirmed by observations. This has motivated extensive research into horizonless ultra-compact objects, including gravastars, rebounding shells, and wormholes, which can closely mimic black holes in their outer regions \citep{cardoso_1}. Such configurations provide an interesting setting for investigating the interplay between classical gravitational dynamics and quantum effects.

A particularly attractive possibility is to describe a spherically symmetric wormhole as a dynamical system whose collective coordinate is the radius of its throat. This reduces the problem of gravitational dynamics to an effectively one-dimensional system and opens the way to its quantum-mechanical treatment.

The quantum mechanics of wormhole throats has a substantial history. An early investigation by \citep{visser2} demonstrated the possibility of quantum stabilization of Minkowski-signature wormholes. More recently, \citep{borissova} developed a quantum description of a Minkowski–Minkowski wormhole, using an effective action for the throat radius derived from the Israel junction conditions. They constructed a path-integral representation of the quantum propagator and investigated the suppression of topology-changing transitions within a semiclassical approximation. In these models, the wormhole throat is a non-degenerate junction surface whose dynamics is associated with a thin-shell configuration. In a related but different approach, \citep{magueijo} investigated the quantum geometry of the Einstein–Rosen bridge using a symmetry-reduced canonical formulation, revealing quantum correlations between the two asymptotically flat regions.

The present work addresses a complementary class of configurations: matter-free degenerate wormholes, whose metric determinant vanishes at the throat. Such geometries originate from the Einstein–Rosen construction \citep{einstein} and have attracted renewed interest following the studies of degenerate wormholes \citep{horowitz, klinkhamer}. Unlike non-degenerate thin-shell wormholes, which require a surface stress-energy tensor, degenerate wormholes can be described as vacuum configurations within the regularized Einstein equations. In particular, a spherically symmetric degenerate wormhole can possess a Schwarzschild exterior characterized by a nonzero gravitational mass parameter \textit{M}, while remaining free of matter sources \citep{dimaschko4}.

The essential difference between these two classes of wormholes concerns the origin of their dynamics. In the thin-shell approach, the motion of the throat is governed by the junction conditions and an associated effective action. For a degenerate matter-free throat, by contrast, the local vacuum field equations do not uniquely determine the evolution of its radius. An additional dynamical principle is therefore required.
 
In \citep{dimaschko2,dimaschko3}, such a principle was formulated in terms of an extended equivalence principle, which identifies the normal motion of a matter-free degenerate throat with the free fall of a test particle in the corresponding gravitational field. This principle was applied to the radial collapse of a spherically symmetric degenerate wormhole and subsequently given a local geometric justification in the planar Rindler sector. For a wormhole with a Schwarzschild exterior, it reduces the radial dynamics of the throat to the standard problem of radial free fall in a Schwarzschild gravitational field.

This correspondence provides the starting point for the present quantum-mechanical investigation. The radial dynamics of the throat can be reduced to a one-dimensional Kepler problem \citep{landau}, which in turn admits an oscillator representation \citep{LC}. The corresponding quantum problem possesses a dynamical symmetry described by the non-compact group \(SU(1,1)\), allowing an algebraic determination of its discrete energy spectrum \citep{dmitriev,barut}.

The principal purpose of the present work is to construct a quantum-mechanical description of the collective radial dynamics of a matter-free degenerate wormhole. In contrast to the path-integral treatment of non-degenerate thin-shell wormholes, the present approach leads to an algebraically solvable spectral problem. We determine the energy spectrum and the corresponding quantum states, and investigate their dependence on the ratio between the wormhole mass \(M\) and the Planck mass \(m_P\).

Two qualitatively different regimes emerge. For \(M<m_P\), the system exhibits a Bohr-type spectrum analogous to the bound-state spectrum of the hydrogen atom. For \(M>m_P\), the spectrum undergoes a qualitative change, leading to a non-Bohr regime associated with the semiclassical Einstein–Rosen configuration. We also examine the observer-dependent interpretation of the quantum states and discuss possible implications for the evolution of the wormhole under a gradual decrease of its gravitational mass.

The paper is organized as follows. Section 2 introduces the spherically symmetric degenerate wormhole geometry used in the subsequent analysis. Section 3 considers the classical dynamics of its throat in the co-moving and distant reference frames and establishes the reduction to a one-dimensional oscillator. Section 4 develops the quantum-mechanical formulation using the dynamical group \(SU(1,1)\), determines the energy spectrum and quantum states, and discusses the distinction between visible and hidden states together with their possible evolution under mass decrease. Section 5 summarizes the principal results and their physical implications.

\section{Schwarzschild-like degenerate wormhole}

In this work, we consider a Schwarzschild-like degenerate wormhole, defined as a spherically symmetric, matter-free, two-sheeted configuration whose exterior geometry on each sheet is described by the Schwarzschild metric. We assume the same mass parameter \(M\) on both sheets, corresponding to a mirror-symmetric configuration.

Here, we derive this configuration from the Schwarzschild solution and briefly discuss its fundamental properties.  

The metric of the wormhole can be obtained from the standard Schwarzschild metric\footnote{The Planck system of units is used, in which the speed of light \(c\), the gravitational constant \(G\), and the Planck constant \(\hbar\) are equal to 1. In this system, the unit of mass is \(m_P=2.18 \cdot10^{-8}\) kg, the unit of length is \(l_P=1.62 \cdot 10^{-35}\)m, the unit of time is \(t_P=5.39 \cdot 10^{-44}\)s J. The metric (1) is written in usual spherical coordinates \((r, \theta, \varphi)\), \( d\Omega^2=d\theta^2+sin^2 \theta   d\varphi^2  \) is an element of solid angle, \(t\) is time, \(M\) is the mass of the source of the gravitational field.}
\begin{equation}
    ds^{2} = -\left(1 - \frac{2M}{r}\right) dt^{2}
    + \left(1 - \frac{2M}{r}\right)^{-1} dr^{2}
    + r^{2} d\Omega^{2}
\end{equation}after applying the topological dressing operation to it \citep{dimaschko,dimaschko4}. To do this, it is necessary to choose a two-sheeted mapping that projects the two-sheeted space of the wormhole onto the original one-sheeted space of the Schwarzschild metric. By choosing the mapping
\begin{equation}
    r = \sqrt{l^2 + b^2},
\end{equation} with constant \(b>0\), we obtain the stationary metric
\begin{equation}
   \begin{split}
 ds^{2}
    = -\left(1 - \frac{2M}{r}\right) dt^{2}
    + \left(1 - \frac{2M}{r}\right)^{-1}
      \frac{l^{2}}{r^{2} } dl^{2}
    + r^{2} d\Omega^{2}.
   \end{split}
\end{equation}The new variable \(l\) covers all the values from \(-\infty\)  to \(+\infty\), where the positive values \(l>0\) correspond to the first sheet, the negative values \(l<0\) correspond to the second sheet of the two-sheeted space and the value \(l=0\) corresponds to the transition surface between them (i.e., the spherical throat \(r=b\)). The constant parameter \(b \ge 2M\) has the meaning of the throat radius.

The metric (3) is degenerate because its determinant
\begin{equation}
    g(x) \equiv \det\left\| g_{\mu\nu}(x) \right\| = -l^{2}\left(l^{2} + b^{2}\right)\sin^{2}\theta
\end{equation}
vanishes at \(l=0\), i.e., at the spherical throat \(r=b\). This metric constitutes a stationary solution of the \(g^2\)-modified vacuum Einstein equations\footnote{We employ the standard notation: \(g_{\mu\nu}\) denotes the metric tensor, while \(R_{\mu\nu}\) and \(R\) represent the Ricci tensor and the Ricci scalar, respectively.}\footnote{For a degenerate metric, Einstein’s field equations must be modified, as originally proposed in \citep{einstein}. The procedure consists of multiplying the equations by \(g^2\), thereby rendering them polynomial in the metric components \(g_{\mu\nu}\). It was furthermore demonstrated in that work that the regularized equations (5) describe a self-sustaining gravitational field that does not require any additional matter or external sources.}
\begin{equation}
    g^{2}  R_{\mu \nu}  = 0.
\end{equation}
However, this does not imply that it describes the stationary state of the wormhole. The reason is that the local metric \(g_{\mu \nu}\) by itself does not furnish a complete description of the underlying spacetime; a full specification is given by the pair \((\mathscr{M}, g_{\mu \nu})\), which, in addition to the local metric field \(g_{\mu \nu}\), encompasses the global geometric and topological properties of the manifold \(\mathscr{M}\). As demonstrated in \citep{dimaschko}, once this fact is taken into account, the application of the least action principle yields simultaneously: (1) Einstein’s field equations for \(g_{\mu \nu}\), and (2) separate equations governing the global geometric and topological characteristics of the manifold \(\mathscr{M}\). For the problem of the radial collapse of a wormhole, the only relevant global characteristic is \(b(t)\), the (time-dependent) radius of the wormhole throat.

Under the assumption of spherical symmetry, the analysis is significantly simplified by  the extended equivalence principle applied to the motion of the throat in its own gravitational field. In this framework  the problem of the radial collapse of the wormhole is effectively mapped onto the standard problem of the radial infall of a test particle in a Schwarzschild spacetime \citep{dimaschko2}. This correspondence was shown to uniquely fix the dynamical evolution equation for \(b(t)\).

\section{Classical wormhole mechanics}
\subsection{Co-moving reference frame}

We will use the methodology presented in \citep{dimaschko2} to describe the collapse of a degenerate throat, which is a spherically symmetric source of a gravitational field. By Birkhoff's theorem \citep{birkhoff}, this field is governed by a Schwarzschild metric. Therefore, in accordance with the extended equivalence principle, the radial dynamics of the throat follows the same equation as the radial fall of a test particle in a Schwarzschild gravitational field.

In this section, we consider the gravitational collapse of the wormhole in the proper time \(\tau\) of an observer located at its throat. The corresponding problem of the radial fall of a test particle is described by equation [\citep{MTW}, Eq.(25.16 a,b))
\begin{equation}
    \left(\frac{dr}{d\tau}\right)^{2} + 1 - \frac{2M}{r} = \frac{\epsilon^{2}}{m^{2}}
    \label{eq:placeholder_label}
\end{equation}Here \(r=r(\tau)\) is the radial coordinate of the particle as a function of the proper time \(\tau\), and \(\epsilon\) and \(m\) are, respectively, the total energy of the particle and its rest mass. The collapse of the wormhole in the proper time \(\tau\) is described by the same equation for a variable radius of the wormhole  \(b(\tau)\):

\begin{equation}
    \left(\frac{db}{d\tau}\right)^{2} + 1 - \frac{2M}{b} = \frac{E^{2}}{M^{2}}.
\end{equation}Here \(E\) and \(M\) are the energy and mass of the collapsing wormhole, respectively. Equation (7) maps the dynamics of the wormhole to the one-dimensional Kepler problem. In this paper, we are only interested in bound states describing gravitational collapse, for which \(E^2<M^2\). In this case, the square of the wormhole energy \(E^2\) can be represented as
\begin{equation}
    E^2 = (1 - \omega^2) M^2,
\end{equation}where the parameter \(\omega\) is limited to the region \(0<\omega<1\). This parameterization reduces equation (7) to a “non-relativistic” form:
\begin{equation}
    \frac{1}{2} \left( \frac{db}{d\tau} \right)^{2} - \frac{M}{b} = -\frac{1}{2} \omega^{2}.
\end{equation}It is a classical result that the two-dimensional Kepler problem can be mapped onto a two-dimensional harmonic oscillator \citep{LC}. The same Levi-Civita regularization procedure can be employed to transform the one-dimensional Kepler problem (9) into an equivalent one-dimensional harmonic oscillator problem. For this purpose, it suffices to replace the proper time \(\tau\) with a new time parameter \(\eta\), defined through the relation
\begin{equation}
    d\eta = \frac{\omega}{b} \, d\tau.
\end{equation}
In what follows, we shall refer to the time parameter \(\eta\) as the oscillator time. Upon reparametrizing the system in terms of the oscillator time \(\eta\), the original equation (9) is transformed into the equation of motion for a unit-frequency harmonic oscillator with total energy \(M^2/(2\omega^4)\):
\begin{equation}
    \frac{1}{2} \left( \frac{db}{d\eta} \right)^{2}
    + \frac{1}{2} \left( b - \frac{M}{\omega^{2}} \right)^{2}
    = \frac{M^{2}}{2 \omega^{4}}.
\end{equation}
The following oscillatory law of motion \(b(\eta)\), satisfying the initial condition \(b(0)=2M/\omega ^2\), has the form
\begin{equation}
    b = \frac{M}{\omega^2} \left( 1 + \cos \eta \right).
\end{equation}Taking into account the differential relation (10), this allows us to express the proper time \(\tau\) in terms of the oscillatory time \(\eta\):
\begin{equation}
    \omega \tau = \frac{M}{\omega^{2}} \left( \eta + \sin \eta \right).
\end{equation}
Parametric equations (12) and (13) determine the dependence of the throat radius \(b\) on the proper time \(\tau\), which is shown in Fig. 1. This dependence is identical to the parametric dependence on the proper time \(\tau\) of the radial coordinate \(r\) of a particle in a collapsing dust cloud in a spherically symmetric gravitational collapse [\citep{landau}, Eq. (103.16)]. In the absence of non-gravitational forces, such a coincidence is a direct consequence of the equivalence principle.
\begin{figure} [t]
 \centering
 \includegraphics[scale=0.5]{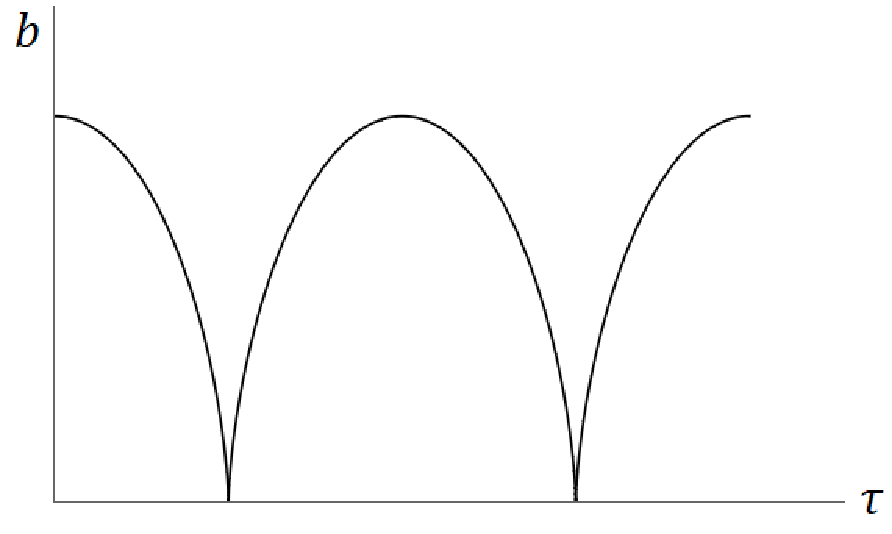}
  \caption{Oscillations of the radius \(b\) of the Schwarzschild-like degenerate wormhole in proper time \(\tau\)}
\end{figure}

The dependence of the wormhole radius \(b\) on the proper time \(\tau\)  is periodic with a period of
\begin{equation}
    T = 2\pi \frac{M}{\omega^{3}}.
\end{equation}It is generally accepted that in the case of spherical collapse of dust matter, such a periodic dependence on the proper time \(\tau \) is a consequence of neglecting pressure \citep{MTW}, as well as possible quantum effects \citep{hawking_ellis}. For the  Schwarzschild-like degenerate wormhole, matter and the associated pressure are absent. In subsection 4.3 of this paper, we show that in the case of a degenerate wormhole under consideration, it is quantum effects that effectively "cut off" this periodic dependence. For values of \(b \sim l_P\) quantum effects come into play, and the classical dependence (12, 13) becomes inapplicable.

\subsection{Distant reference frame}

In this section, we consider the same problem of the gravitational collapse of a wormhole, now not in the \textit{co-moving} frame of reference, but from the perspective of a \textit{distant observer}. To do this, we must move from proper time \(\tau\) to absolute time \(t\), which, in the case of a test particle, is realized by the relation [\citep{MTW}, Eq. (25.18)]:
\begin{equation}
    \frac{\epsilon}{m} \, d\tau = \left(1 - \frac{2M}{r}\right) dt
\end{equation} Applying the equivalence principle, in this relationship we replace the radial coordinate \(r\), energy \(\epsilon\), and rest mass \(m\) of the test particle, respectively, by the radius \(b\), energy \(E\) and mass \(M\) of the wormhole:

\begin{equation}
    \frac{E}{M} \, d\tau = \left(1 - \frac{2M}{b}\right) dt.
\end{equation}The meaning of this relationship remains unchanged and consists of taking into account time dilation near the source of the gravitational field. Substituting here for \(b\) and \(\tau\) their expressions according to (12) and (13), and then integrating over \(d \eta\), we obtain an explicit expression for the absolute time t in terms of the oscillatory time \(\eta\):

\begin{equation}
   \begin{split}
\frac{t}{M}
   & = \sqrt{\frac{1}{\omega^{2}} - 1}
      \left[
        \left(\frac{1}{\omega^{2}} + 2\right)\eta
        + \frac{1}{\omega^{2}} \sin \eta
      \right] \\ 
   &   \quad  + 4 \tanh^{-1}
      \left[
        \tan \left(\frac{\eta}{2} \right) \left (\sqrt{\frac{1}{\omega^{2}} - 1}  \,\,\right )^{-1}
      \right]
   \end{split}
\end{equation}In accordance with the equivalence principle, it coincides with a similar expression for a radially falling particle in the Schwarzschild gravitational field [\citep{MTW}, Eq. (25.37)]. Together with the previous equation (12), this expression parametrically determines the evolution of the radius \(b\) of the wormhole in absolute time \(t\), i.e., from the point of view of a \textit{distant observer}. Note that expression (17) remains real only up to the boundary value  \(\eta =\bar{\eta}\), when the argument of the inverse hyperbolic tangent becomes equal to unity:
\begin{equation}
    \tan\left(\frac{\bar{\eta}}{2}\right) = \sqrt{\frac{1}{\omega^{2}} - 1}.
\end{equation}In oscillatory time, \(\bar{\eta}\) corresponds to the moment at which communication between the \textit{co-moving} and \textit{distant} observers is lost. It follows from expression (12) that it is precisely at this moment \(\eta =\bar{\eta}\) that the wormhole's throat intersects the horizon, i.e. \(b=2M\). 

As in the case of a radially falling test particle, gravitational time dilation stretches the absolute time scale \(t\) as compared to the proper time scale \(\tau\) and transforms the initially periodic, cycloidal dependence of the wormhole radius on time (see Fig. 1) into a monotonic one. This monotonically decreasing dependence, tending to the limiting value \(b=2M\) at infinity, is shown in Fig. 2. It demonstrates that in the limit \(t \rightarrow \infty\), the wormhole asymptotically collapses into an Einstein-Rosen wormhole (see also \cite{dimaschko2}).

\begin{figure} [t]
 \centering
 \includegraphics[scale=0.43]{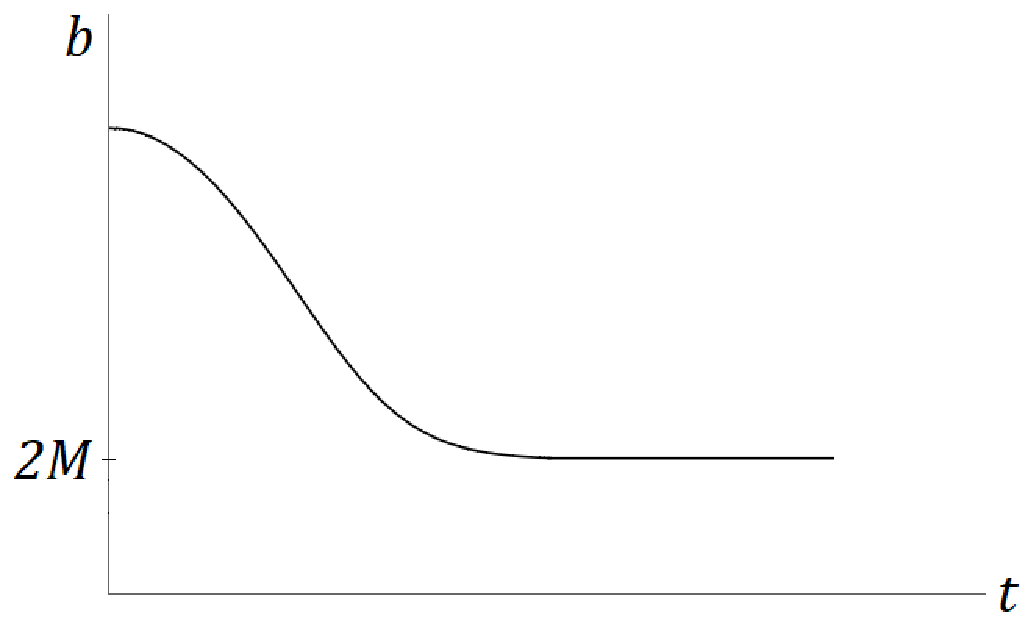}
  \caption{Reduction of the radius \(b\) of the Schwarzschild-like degenerate wormhole in absolute time \(t\).}
\end{figure}
Although the function \(b(t)\) appears non-oscillatory in the time coordinate \(t\), it remains oscillatory when expressed in affine oscillatory time \(\eta\) as given by equation (12).  The inability to see this is due solely to the position of the \textit{infinitely distant observer} and the choice of the reference frame corresponding to this position.

Summarizing, the classical (non-quantum) collapse of the Schwarzschild-like degenerate wormhole is completely analogous to the radial infall of a test particle in a Schwarzschild gravitational field, consistent with the equivalence principle. In a \textit{co-moving} frame this collapse appears as cycloidal oscillations, while for a \textit{distant observer} the wormhole asymptotically becomes a non-traversable Einstein–Rosen wormhole. 

\section{Quantum wormhole mechanics}
\subsection{Co-moving reference frame }

Let us move on to the quantum description of the wormhole, where the only "quantum" variable is the radius \(b\) of its throat. In this section, we will consider the quantum behavior of the wormhole in the \textit{co-moving} frame of reference. 

Before proceeding to quantization, we specify the canonical formulation underlying the classical equation of motion (7). Unlike the thin-shell approach, where an effective Lagrangian is obtained from the Einstein–Hilbert action \citep{visser2,borissova}, the present dynamics is determined by the extended equivalence principle \citep{dimaschko2,dimaschko3}. An effective Lagrangian reproducing this dynamics can be chosen as
\begin{equation}
  L(b,\dot b)=\frac{M}{2}\dot b^2+\frac{M^2}{b},
\qquad \dot b=\frac{db}{d\tau}.
\end{equation}Here, the coefficient \(M\) of the kinetic term reflects the identification of the inertial and gravitational mass of the wormhole. The canonical momentum and the corresponding Hamiltonian are, therefore,
\begin{equation}
  p=\frac{\partial L}{\partial\dot b}
=M\dot b,
\qquad
H=\frac{p^2}{2M}-\frac{M^2}{b}.
\end{equation}
According to Eq. (7), the conserved quantity \(E^2\) is related to this Hamiltonian by \(E^2=M^2+2MH\). Canonical quantization of this relation yields the spectral operator  
\begin{equation}
    \hat{H} = \hat{p}^{2} + M^{2}\left(1 - 2M \hat{b}^{-1}\right)
\end{equation}which enters the Schrödinger equation
\begin{equation}
    (\hat{H} - E^{2}) \lvert \psi \rangle = 0.
\end{equation}
Here, \(\hat{p}=-i \partial / \partial b\)  is the canonical momentum operator corresponding to the classical momentum \( p=M db/d \tau\) with \([\hat{b},\hat{p}]=i\).  In the following, we consider the bound-state sector using the same form of parameterization \(E^2=(1-\omega^2)M^2\) as in the classical description, Eq.(8).

To diagonalize the one-dimensional Hamiltonian \(\hat{H}\), we employ the dynamical group method based on the non-compact Lie group \(SU(1,1)\). This approach was originally formulated by Dmitriev and Rumer in the context of the hydrogen atom \citep{dmitriev} and  by Barut and Girardello to the Kepler problem \citep{barut}.

According to this method, the Schrödinger equation (22) should be reduced to the action on the quantum state \(\lvert \psi \rangle\) of a linear combination of generators of the group \(SU(1,1)\). To achieve this, we left-multiply the Schrödinger equation by the operator \(\hat{b}\), upon which it assumes the form
\begin{equation}
    \left(\hat{b}\hat{p}^{2} - 2M^{3} + \hat{b}\omega^{2}M^{2}\right ) \, \lvert \psi \rangle = 0.
\end{equation}
We now introduce the following set of three operators:
\begin{equation}
    \hat{K}_0 = \frac{1}{2}\bigl(\hat{b}\hat{p}^2 + \hat{b}\bigr),
\end{equation}
\begin{equation}
    \hat{K}_1 = \frac{1}{2}\bigl(\hat{b}\hat{p}^2 - \hat{b}\bigr),
\end{equation}
\begin{equation}
    \hat{K}_2 = \hat{b}\hat{p},
\end{equation}
which satisfy the commutation relations,
\begin{equation}
    [\hat{K}_1,\hat{K}_2] = -i\hat{K}_0, \quad
    [\hat{K}_2,\hat{K}_0] = i\hat{K}_1, \quad
    [\hat{K}_0,\hat{K}_1] = i\hat{K}_2,
\end{equation}
and thereby span a representation of the Lie algebra \(\mathfrak{su}(1,1)\), corresponding to the group \(SU(1,1)\) \citep{perelomov}. The associated Casimir operator \(\hat{C}\) takes the standard form
\begin{equation}
    \hat{C} = \hat{K}_0^2 - \hat{K}_1^2 - \hat{K}_2^2 = \phi (\phi + 1)\,\hat{I},
\end{equation}
where \(\hat{I}\) denotes the identity operator and, for unitary representations, the parameter satisfies \(\phi < 0\). The value of \(\phi\) uniquely specifies the discrete spectrum of the operator \(\hat{K}_0\):
\begin{equation}
    \hat{K}_0 \lvert n \rangle = (-\phi + n)\lvert n \rangle, \quad  n = 0,1,2,\dots .
\end{equation}In this representation, \(\hat{C}=0\), which implies \(\phi=-1\). Consequently, the discrete spectrum of the operator \(\hat{K}_0\) commences at unity and is given by
\begin{equation}
    \hat{K}_0 \lvert n \rangle = n \lvert n \rangle, \quad   n = 1,2, 3\dots  .
\end{equation}Linear transformations \(\hat{U}\) belonging to the group \(SU(1,1)\), when applied to an arbitrary linear combination of the generators defined by the pseudo-Euclidean scalar product
\begin{equation}
    (\vec{h}, \hat{\vec{K}}) = h_0 \hat{K}_0 - h_1 \hat{K}_1 - h_2 \hat{K}_2,
\end{equation}
preserve the associated pseudo-Euclidean norm
\begin{equation}
    \left|\vec{h}\right| \equiv \sqrt{h_0^2 - h_1^2 - h_2^2}.
\end{equation}
Next, we exploit the fact that the operator Schrödinger equation in the form (23) is linear in the generators,
\begin{equation}
    \bigl[ \left(\hat{K}_0 + \hat{K}_1\right) - 2M^3 + \omega^2 M^2 \left(\hat{K}_0 - \hat{K}_1\right) \bigr] \lvert \psi \rangle = 0,
\end{equation}
so that it can be associated with the vector
\begin{equation}
    \vec{h} = \left( \omega^2 M^2 + 1,\ \omega^2 M^2 - 1,\ 0 \right).
\end{equation}
In accordance with Eq. (32), the norm of this vector is given by
\begin{equation}
    \left\lvert \vec{h} \right\rvert = 2 \omega M.
\end{equation}
Consequently, for an appropriate choice of the group transformation \(\hat{U}\), the Schrödinger operator in Eq.(33) can be expressed solely in terms of the transformed operator \(\hat{K}_0\),
\begin{equation}
    \hat{U}\,\left(2\omega M \hat{K}_0 - 2M^3\right)\,\hat{U}^{-1}\,\lvert \psi \rangle = 0.
\end{equation}A suitable diagonalizing group operator is given by the rotation
\begin{equation}
    \hat{U} = \exp\left(-i \theta \hat{K}_2\right),
\end{equation}
which implements a linear transformation in the \((0 1)\) plane:
\begin{equation}
    \hat{U}\,\hat{K}_0\,\hat{U}^{-1} = \hat{K}_0 \cosh\theta + \hat{K}_1 \sinh\theta,
\end{equation}
\begin{equation}
    \hat{U}\,\hat{K}_{1}\,\hat{U}^{-1} = \hat{K}_{0}\,\sinh\theta + \hat{K}_{1}\,\cosh\theta.
\end{equation}The rotation angle \(\theta\) required to achieve diagonalization is given by
\begin{equation}
    \theta = \ln\left(\frac{1}{M\omega}\right).
\end{equation}
The diagonal form (36) of the Schrödinger equation, together with the known spectrum (28) of the generator \(\hat{K}_{0}\), fixes the allowed values of the parameter \(\omega\) as
\begin{equation}
    \omega = \omega_n = \frac{M^2}{n}, \quad   n = 1, 2, 3, \dots  ,
\end{equation}
and, in accordance with relation (8), leads to the discrete energy spectrum of a wormhole with mass \(M\):
\begin{equation}
    E = E_n = M \sqrt{1 - \frac{M^4}{n^2}}, \quad  n = 1, 2, 3, \ldots   .
\end{equation}
The corresponding stationary quantum states of the wormhole are as follows.
\begin{equation}
   \lvert \psi_n \rangle = \hat{U}\lvert n \rangle .
\end{equation}
In particular, for the ground state (\(n=1\)) one finds
\begin{equation}
    E_1 = M\sqrt{1 - M^4},
\end{equation}
and the associated quantum ground state of the wormhole is
\begin{equation}
   \lvert \psi_1 \rangle = \hat{U}\lvert 1 \rangle .
\end{equation}
Since the radicand in the general expression (42) for the energy spectrum must be nonnegative, one obtains the inequality
\begin{equation}
    n \geq M^2,
\end{equation}
which imposes a lower bound on the admissible values of the quantum number \(n\).

In particular, for \(M \le 1\), the wormhole admits a complete discrete spectrum \(n = 1, 2, 3, \ldots\), which is analogous to the Bohr energy spectrum of the hydrogen atom. Accordingly, a wormhole with sufficiently small mass, \(M \le 1\), will be referred to as a \textit{Bohr wormhole}. In conventional (non Planck) units, the condition \(M \le 1\) corresponds to a wormhole whose mass is less than the Planck mass \(m_P\). Fig. 3 illustrates the spectrum of a \textit{Bohr wormhole} with mass \(M = 0.7\). In accordance with (40) and (44), this spectrum includes all possible energy levels \(n = 1, 2, 3, \ldots\), which are depicted by solid lines.

\begin{figure} [t]
 \centering
 \includegraphics[scale=0.35]{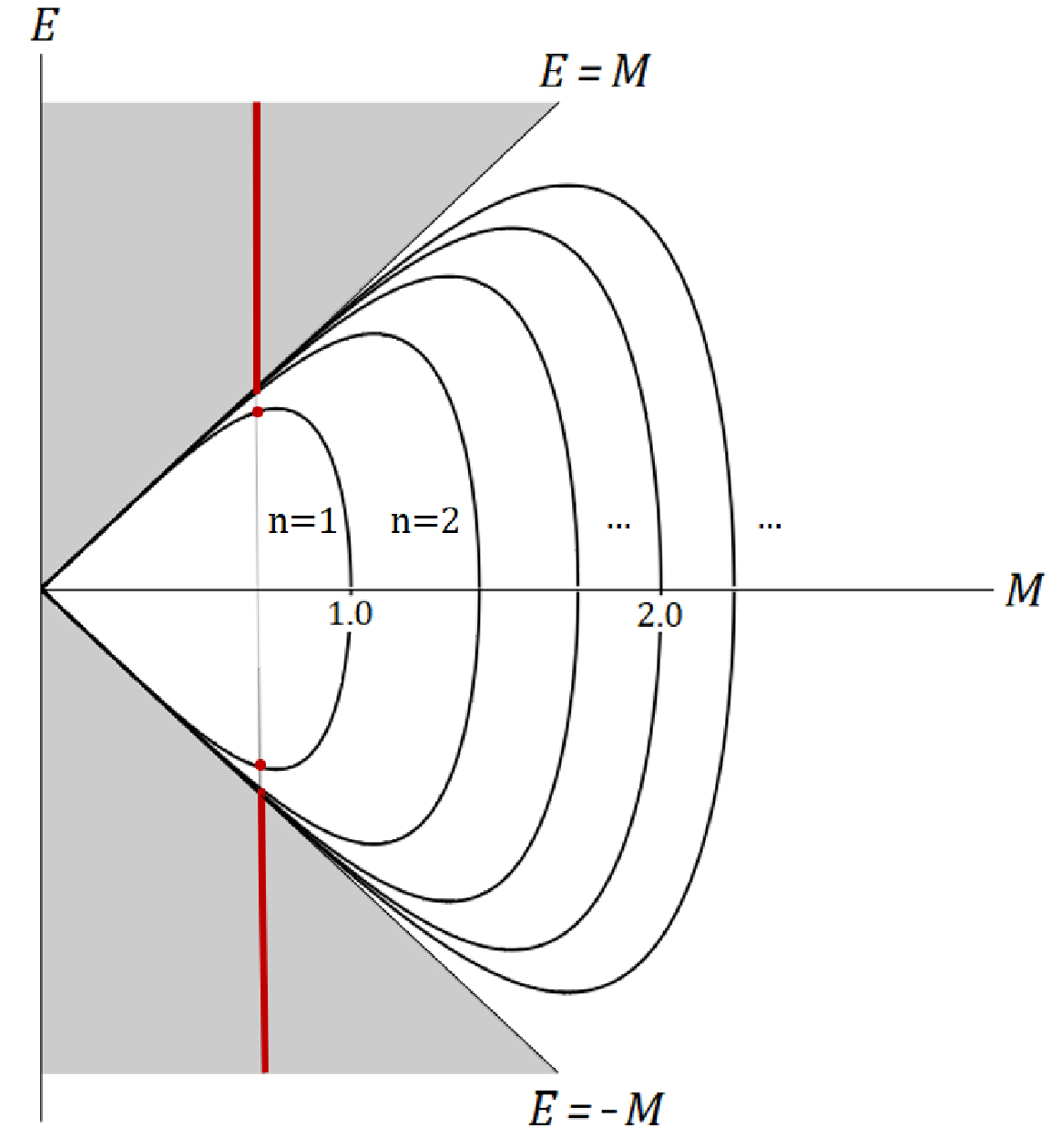}
\caption{Energy spectrum of a wormhole as a function of its mass \(M\). The spectrum of a \textit{Bohr-type wormhole} with \(M<1\) (specifically \(M=0.7\)) is shown in red. It displays a complete set of discrete energy eigenvalues, characterized by the quantum number \(n=1,2,3,\ldots\). The continuous components of the spectrum for \(E>M\) and \(E<-M\) are represented in gray.}
\end{figure}

\begin{figure} [h]
 \centering
 \includegraphics[scale=0.35]{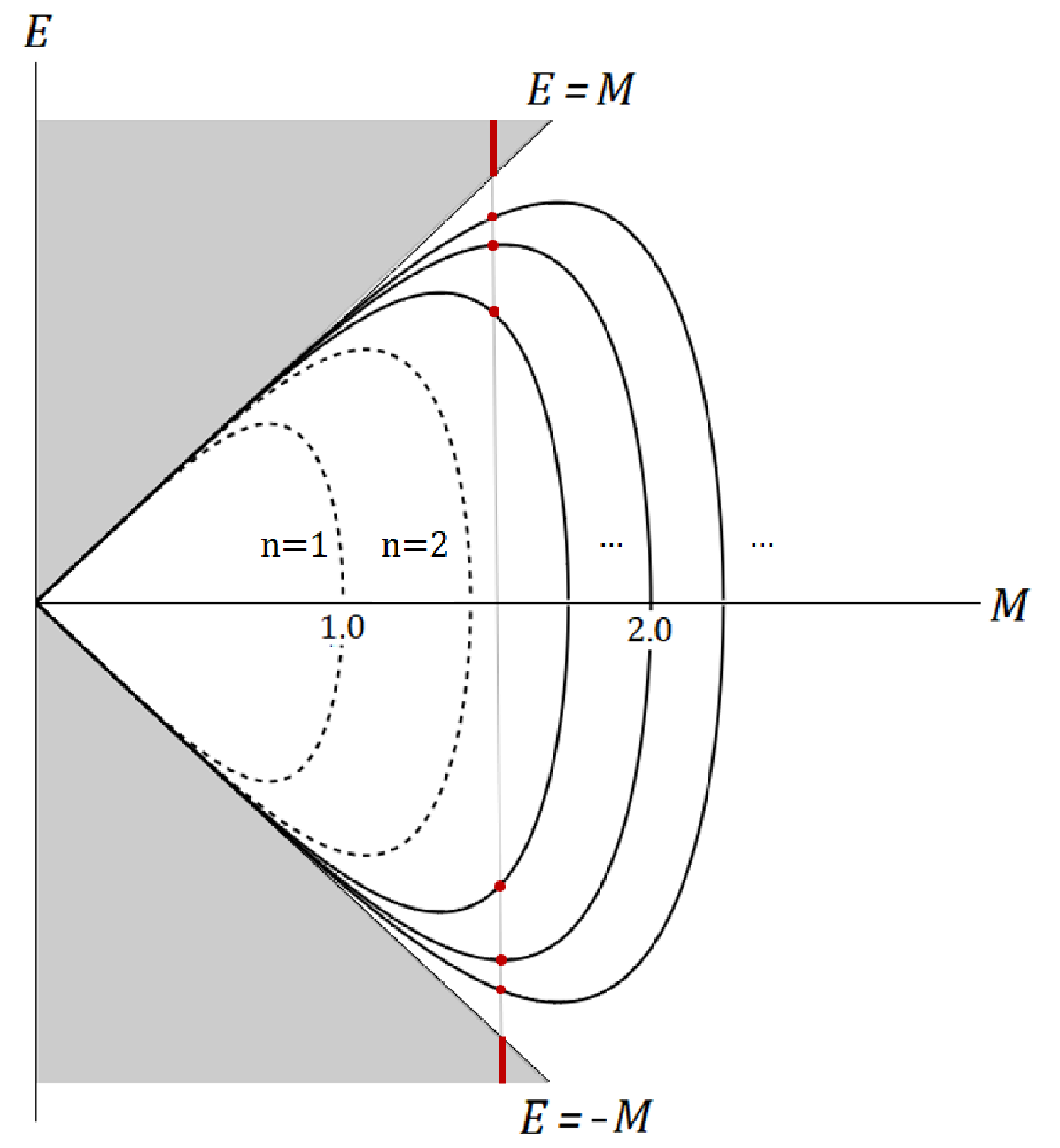}
\caption{Energy spectrum of a wormhole as a function of its mass \(M\). The spectrum of a \textit{non-Bohr wormhole}, \(M>1\) (here \(M=1.5\)), is shown in red. The spectrum contains discrete levels \(n=3,4,5, ...\) . The missing lower levels \(n=1,2\) are shown by the dashed line. The continuum \(E>M\) and \(E<-M\) is shown in gray.}
\end{figure}
As the dimensionless mass parameter \(M\) increases beyond the critical value \(M = 1\), the lowest-lying energy levels of the wormhole successively disappear. In accordance with constraint (46), this occurs at the discrete mass values
\begin{equation}
    M = M_n = \sqrt{n}, \quad n = 1, 2, 3, \ldots .
\end{equation}
Consequently, a wormhole with sufficiently large mass \(M > 1\), whose Bohr-type spectrum is therefore incomplete, will be referred to as a \textit{non-Bohr wormhole}. In conventional physical units, condition \(M > 1\) corresponds to a wormhole whose mass exceeds Planck mass \(m_P\). Fig. 4 presents the energy spectrum of a \textit{non-Bohr wormhole} with mass \(M = 1.5\). According to constraint (46), this spectrum lacks the two lowest levels, \(n = 1\) and \(n = 2\), which are indicated by the dashed line.

In addition to the quantum energy spectrum, the quantum expectation value of the wormhole radius is of interest. Since the wormhole radius operator \(\hat{b}\) is given as a linear combination of algebra generators,
\begin{equation}
    \hat{b} = \hat{K}_0 - \hat{K}_1,
\end{equation}
the quantum expectation value of the radius in the \(n\)-th stationary state \(\lvert \psi_n \rangle\) can be straightforwardly evaluated:
\begin{equation}
    b_n \equiv \langle \psi_n \lvert \hat{b} \rvert \psi_n \rangle 
    = \langle n \lvert \hat{U}^{-1} \hat{b} \hat{U} \rvert n \rangle 
    = n e^{\theta}.
\end{equation}Taking into account Refs. (40, 41), this leads to the following expression for the average throat radius:
\begin{equation}
    b_n = \frac{n^2}{M^3}\equiv n^2 b_B,
\end{equation}
where
\begin{equation}
    b_B  \equiv \frac{1}{M^3}
\end{equation}
defines the so-called “\textit{gravitational Bohr radius}”. 

\subsection{Distant reference frame}
In this section, we revisit the quantum description of a collapsing wormhole, now formulated from the perspective of a \textit{distant observer} rather than in the \textit{co-moving} reference frame. To this end, we must transform the proper time \(\tau\) to the coordinate (asymptotic) time \(t\) using the classical relation (16). This relation can equivalently be expressed as
\begin{equation}
    \frac{b}{M} \frac{db}{d\tau} = (b - 2M)\,\frac{1}{E}\,\frac{db}{dt},
\end{equation}
whose quantum counterpart is implemented via the corresponding operator substitution
\begin{equation}
    \hat{b}\hat{p} \;\rightarrow\; (\hat{b} - 2M)\hat{p},
\end{equation}
or, more succinctly, \(\hat{b} \rightarrow \hat{b}-2M\). This shift is generated by the unitary operator
\begin{equation}
    \hat{V} = \exp\left(-2iM\hat{p}\right),
\end{equation}
which induces the necessary transformation
\begin{equation}
    \hat{b} \;\rightarrow\; \hat{V}\,\hat{b}\,\hat{V}^{-1} = \hat{b} - 2M.
\end{equation} The Hamiltonian (20) must also undergo the same transformation:
\begin{equation}
    \hat{H} \rightarrow \hat{V} \hat{H} \hat{V}^{-1}.
\end{equation}This means that within the quantum-mechanical framework, the classical substitution of time given by (16) must be replaced by the unitary transformation \(\hat{V}\) as shown  in (54)–(56)\footnote{Strictly speaking, the co-moving and distant descriptions are defined in different Hilbert spaces, \(\mathcal H_{\mathrm{co}}\) and \(\mathcal H_{\mathrm{dist}}\), respectively. The operator \(V:\mathcal H_{\mathrm{co}}\to\mathcal H_{\mathrm{dist}}\) implements the shift of the radial coordinate \(b\to\bar b=b+2M\) and is understood as a unitary mapping between the corresponding representations, rather than as a unitary translation within a single Hilbert space on the positive half-line \(b \ge 0\). The analogous coordinate shift \(r\to\bar r=r+2M\) in the Schwarzschild geometry is known as the Brillouin transformation \citep{brillouin}.}.

Thus, we possess a unitary operator \(\hat{V}\) that implements the transformation from the \textit{co-moving} reference frame to the reference frame of a \textit{distant observer}. By acting with this unitary operator on the complete set of quantum states \(\{\lvert \psi_n \rangle\}\), defined in the \textit{co-moving} frame, we obtain a corresponding complete set of quantum states \(\{\lvert \bar{\psi}_n \rangle\}\) in the frame of the \textit{distant observer}:
\begin{equation}
    \lvert \psi_n \rangle \;\rightarrow\; \lvert \bar{\psi}_n \rangle = \hat{V} \lvert \psi_n \rangle .
\end{equation}
The simultaneous unitary transformation of the Hamiltonian and its corresponding eigenstates, as prescribed by Eqs. (55) and (56), preserves the eigenvalue spectrum. Therefore, the Hamiltonian spectrum obtained in the frame of the \textit{co-moving} observer must be identical to that measured by the \textit{distant observer}.

Conversely, the eigenstates \(\{\lvert \psi_n \rangle\}\) defined in the \textit{co-moving} frame of reference and the eigenstates \(\{\lvert \bar{\psi}_n \rangle\}\) defined in the \textit{distan}t frame of reference are, naturally, distinct. In particular, this distinction is reflected in the fact that the expectation value of the wormhole radius, denoted by \(b_n\) in the \textit{co-moving} frame, and the corresponding expectation value \(\bar{b}_n\) in the \textit{distant} frame, do not coincide:
\begin{equation}
   \begin{split}
\bar{b}_n  & \equiv \langle \bar{\psi}_n \lvert \hat{b} \rvert \bar{\psi}_n \rangle  = \langle \psi_n \lvert \hat{V}^{-1} \hat{b} \hat{V} \rvert \psi_n \rangle \\ 
   &  = \langle \psi_n \lvert (\hat{b} + 2M) \rvert \psi_n \rangle
    = b_n + 2M
   \end{split}
\end{equation}
Taking into account Eqs. (50) and (51), one obtains
\begin{equation}
    \bar{b}_n = 2M +\frac{ n^2}{M^3}.
\end{equation}
This expression is in agreement with the corresponding classical description: in the \textit{distant} frame of reference, the wormhole radius \(b\) cannot be smaller than the gravitational radius \(2M\), whereas in the \textit{co-moving} frame of reference no such lower bound is imposed.

\subsection{Visible and hidden states}

We now clarify the physical meaning of the successive loss of the lowest levels from the wormhole spectrum as its mass \(M\) increases. The key observation is that the spectral parameter \(E^2\) appearing in the Schrödinger equation (22) is not constrained to be positive.

From Eq. (58), the eigenvalues \(E_n^2\) of the Hamiltonian  \(\hat{H}\)  are
\begin{equation}
    E^2 = E_n^2 = M^2 \left(1 - \frac{M^4}{n^2}\right), \quad (n = 1, 2, 3, \dots),
\end{equation}
and can have either sign:  \(E_n^2>0\)  for  \(n > M^2\),  \(E_n^2<0\)  for  \(n < M^2\).

It is important to emphasize that \(E_n^2\) here is the constant of motion associated with the throat dynamics in the \textit{co‑moving} (non‑inertial) frame, cf. Eq. (8), rather than the square of a conserved energy defined with respect to an asymptotic time coordinate. In the \textit{co‑moving} description, there is no fundamental obstruction to  \(E^2\) taking negative values: classically, Eq. (8) remains a well‑posed first integral, and quantum‑mechanically, the operator \(\hat{H}\) is self‑adjoint with a discrete spectrum bounded from below in the Hilbert space of bound states. Thus, in the \textit{co‑moving} frame all eigenstates \(\lvert \psi \rangle\) that solve
\begin{equation}
    \left( \hat{H} - E^{2} \right) \lvert \psi \rangle = 0,
\end{equation}are physically admissible, irrespective of the sign of  \(E^2\). By the spectral theorem \citep{von_neumann,griffiths}, the set  \(\lvert \psi_n \rangle\) is \textit{complete} within this bound‑state subspace.

The subtlety arises when we relate the \textit{co‑moving} description to that of an \textit{infinitely distant} (inertial) observer. Classically, this relation involves changing from proper time \(\tau\) to the asymptotic time \(t\) using Eq. (52),  which is only meaningful for real, nonzero \(E\), i.e. for \(E^2>0\). In other words, only trajectories with positive \(E^2\) can be reparametrized into orbits defined with respect to the distant observer’s time \(t\).

At the quantum level, the transformation to the \textit{distant} frame is implemented by the unitary operator \(\hat{V} = \exp\left(-2 i M \hat{p}\right)\) which shifts the throat radius operator  \(\hat{b}\) according to (55) and similarly the Hamiltonian \(\hat{H}\) according to (56). For each eigenstate  \(\lvert \psi_n \rangle\) in the \textit{co‑moving} frame, we thus obtain a corresponding state in the \textit{distant} frame, \(\lvert \bar{\psi}_n \rangle = \hat{V} \lvert \psi_n \rangle\). Because  \(\hat{V}\) is unitary, \(\left\{ \lvert \bar{\psi}_n \rangle \right\}\) is again a \textit{complete} orthonormal set, and the spectrum of  \(\hat{H}\) is unchanged.

Nevertheless, the classical restriction to \(E^2>0\) for a consistent definition of asymptotic time implies that, from the standpoint of a \textit{distant observer}, only those states with \(E_n^2>0\) can be associated with physical processes that unfold in finite \(t\). This motivates the following classification of the states \(\lvert \psi_n \rangle\) in the distant frame:

\begin{itemize}
    \item \textbf{Visible states} \( \lvert \bar{\psi}_n ^+\rangle\): states with \(E_n^2>0\). These correspond to configurations whose dynamics can be parametrized by the asymptotic time \(t\) and are, in principle, accessible to measurement by a \textit{distant observer}.
    \item \textbf{Hidden states} \( \lvert \bar{\psi}_n ^-\rangle\): states with \(E_n^2<0\). These remain perfectly well‑defined quantum states in the \textit{co‑moving} frame but admit no interpretation in terms of real asymptotic time evolution. For a \textit{distant observer}, they correspond to configurations that cannot be reached in any finite \(t\), i.e. they reside effectively “behind the horizon’’ and are not directly observable.
\end{itemize}

In this language, the inequality \( n \ge  M^2\)  acquires a precise meaning. As the mass \(M\) increases and passes through the discrete values \(M = \sqrt{n}\), the corresponding level \(n\) reaches \(E_n^2=0\) and then changes sign. At these threshold masses, the associated state crosses the visibility boundary: a formerly \textit{visible} level with \(E_n^2>0\) becomes a \textit{hidden} level with \(E_n^2<0\). In the spectrum as seen by a \textit{distant observer}, the level effectively “disappears’’ from the set of \textit{visible} states, even though it continues to exist as a bona fide eigenstate in the \textit{co‑moving} frame.

Thus, the successive loss of the lowest levels with increasing \(M\) should not be understood as a dynamical instability or an actual annihilation of quantum states. Rather, it reflects a \textbf{frame‑dependent separation} of the full Hilbert space into a \textit{visible} subsector \(E_n^2>0\) and a \textit{hidden} subsector \(E_n^2<0\), reminiscent of the complementarity between infalling and asymptotic observers in black‑hole physics \citep{susskind}. Any fully quantum description of processes involving such wormholes must in principle account for both sectors, even though a \textit{distant observer} can only probe the \textit{visible} part of the spectrum.

\subsection{Mass evolution}

The classification of the quantum states of a wormhole into \textit{visible} and \textit{hidden} states can be conveniently visualized by plotting the dependence of the average wormhole radius \(\bar{b}_n\) on its mass \(M\) and the quantum number \(n\). According to expression (59), this dependence is represented by a family of curves of the form \(\bar{b}_n = 2M + n^2/M^3\); see Fig. 5. All such curves lie above the line \(\bar{b}_n = 2M\), which corresponds to the gravitational radius of the wormhole.

\begin{figure} [h]
 \centering
 \includegraphics[scale=0.4]{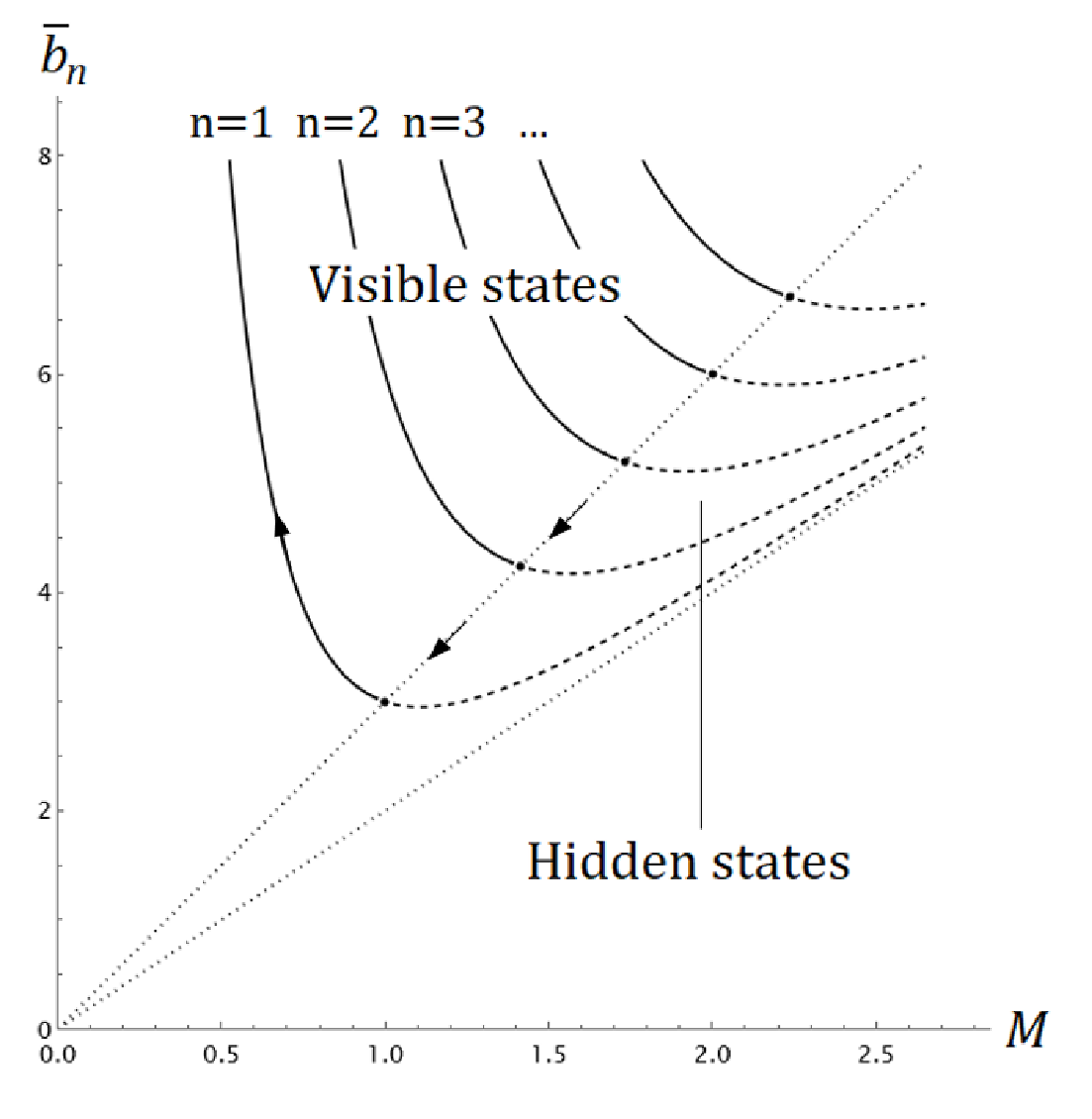}
\caption{Average value \(\bar{b}_n\) of the wormhole radius in a distant reference frame as a function of the mass \(M\) and the quantum number \(n\). The lines \(\bar{b}_n = 2M\) and \(\bar{b}_n = 3M\)   (shown as two dotted lines) bound the region of hidden states (\(2M < \bar{b}_n < 3M\)). The visible states of the wormhole \( \lvert \bar{\psi}_n^+ \rangle\) are shown by solid lines, the hidden states \( \lvert \bar{\psi}_n^- \rangle\)  are shown by their extensions \textit{–} dashed lines. The arrows indicate the direction of evolution of the wormhole radius with decreasing mass \(M\).}
\end{figure}

The transition from \textit{visible} to \textit{hidden} states, which takes place as the mass \(M\) increases, is governed by the spectral parameter \(E^2\) crossing zero. From expression (60)  it follows that the boundary of this transition, determined by the condition \(E^2 = 0\), is represented by the line \(\bar{b}_n = 3M\), corresponding to the photon sphere. The two lines \(\bar{b}_n = 2M\) and \(\bar{b}_n = 3M\) thus enclose the region associated with \textit{hidden} states; both boundaries are indicated by dotted lines in the figure. The plots of the quantum expectation values \(\bar{b}_n\) are displayed for both \textit{visible} states \( \lvert \bar{\psi}_n^+ \rangle\) (solid curves) and \textit{hidden} states \( \lvert \bar{\psi}_n^- \rangle\) (dashed curves).

Invoking the possibility of Hawking evaporation of the wormhole \citep{hawking}, which drives a progressive \textit{decrease} of its mass \(M\), we analyze the associated evolution of the wormhole radius.\footnote{We assume that the mass \(M\) evolves quasi-statically through a sequence of equilibrium configurations, neglecting backreaction effects whose consistent treatment would require a fully dynamical analysis.}  

In the regime \(M>1\) (corresponding to a large, \textit{non-Bohr wormhole}), the reduction of the mass leads to the sequential emergence of new observable states with progressively smaller values of the quantum number \(n\) from beneath the horizon. These newly emergent states successively become the ground state of the system. In Fig. 5, this process is represented by motion along the straight line \(\bar{b}_n = 3M\) toward decreasing values of \(M\), i.e., by a linear decrease of the wormhole radius. At the instant when the lowest-lying level \(n=1\) emerges from below the horizon, the wormhole radius reaches its minimum value \(\bar{b}_n = 3\), which in conventional units is \(3l_P\), i.e., three Planck lengths. This occurs at \(M=1\), or, in standard units, when the wormhole mass \(M\) equals the Planck mass \(m_P\). At this juncture, the \textit{non-Bohr wormhole} is converted into a \textit{Bohr wormhole}, and its ground state becomes fully analogous to the ground state of the hydrogen atom (see Fig. 6).
\begin{figure} [t]
 \centering
 \includegraphics[scale=0.35]{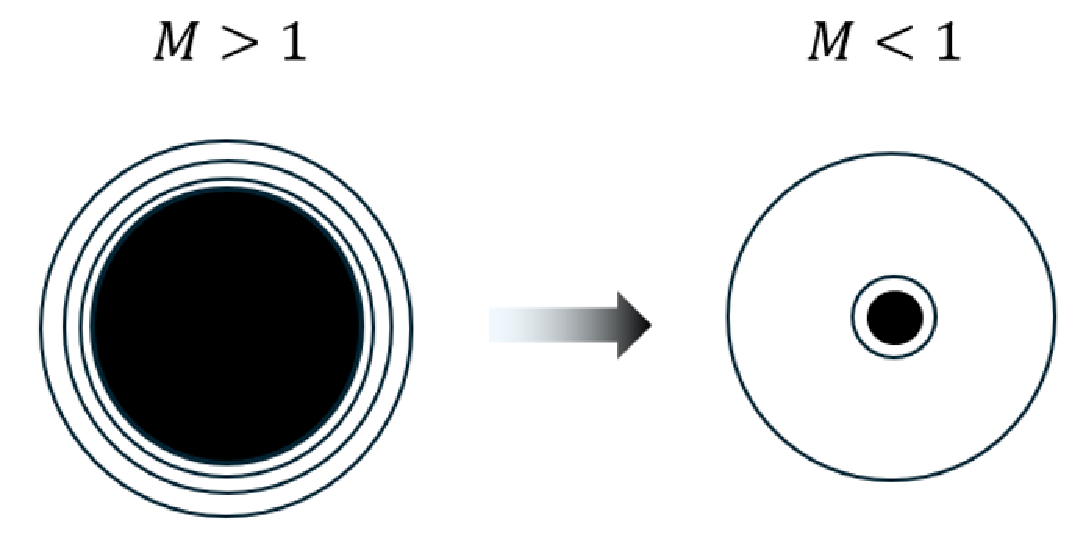}
\caption{Transformation of a \textit{non-Bohr wormhole} (\(M>1\)) into a \textit{Bohr wormhole} (\(M<1\)) as its mass \(M\) decreases due to evaporation. In the region of \(M>1\), the wormhole behaves like an Einstein-Rosen wormhole; in the region of \(M<1\), the wormhole is the gravitational analogue of a hydrogen atom. The transition boundary of \(M=1\) in ordinary units is the Planck mass \(m_P\).}
\end{figure}
As is evident from Fig. 5, a further decrease in the mass \(M\) is expected to result in a rapid increase in the wormhole radius. The critical issue in this context is the efficiency of the Hawking evaporation mechanism: if quantum fluctuations of the gravitational field effectively suppress Hawking radiation at such small wormhole radii \citep{hawking}, then the Planckian configuration with mass \(M = 1\) (the Planck mass \(m_P\)) should remain stable and persist indefinitely. A definitive answer to this question requires a detailed analysis of Hawking evaporation in the framework of the degenerate metric, which lies beyond the scope of the present work.

Another important issue is the intrinsic limitation of the model, namely the assumption of spherical symmetry in the wormhole metric (3). On the one hand, this assumption renders the problem exactly solvable by enabling its reduction to an effectively one-dimensional quantum mechanical system. On the other hand, the same restriction precludes the incorporation of quantum fluctuations that do not respect this symmetry.

It is reasonable to posit that this effect becomes increasingly significant for semiclassical wormholes of larger mass \(M \gg 1\) and, correspondingly, larger areal radius. This constraint exerts a pronounced influence on the localization of the wormhole throat, as illustrated in Fig. 5. In the regime of large masses (\(M > 1\)), the quantum-mechanical expectation value of the wormhole radius lies on the curve \(\bar{b}_n = 3M\), i.e., at the photon sphere, rather than near the gravitational radius \(\bar{b}_n = 2M\), as one might have anticipated. 

The apparent localization at the photon sphere \(\bar{b}_n = 3M\) may encode a genuine physical property of the system: this is precisely the radius at which null geodesics become marginally bound, and thus where quantum vacuum fluctuations of the spacetime geometry are expected to be maximally pronounced. Determining whether this behavior signifies a breakdown or limitation of the present model, or instead reveals a robust physical effect, will require an analysis that goes beyond the assumption of exact spherical symmetry.

On the other hand, the applicability of the spherically symmetric metric (3) for small masses \(M\le 1\) should not raise such doubts.

\section{Summary and Conclusions}

In this work, we have investigated the classical and quantum dynamics of a Schwarzschild-like degenerate wormhole, treating the radius of its matter-free throat as a collective dynamical degree of freedom. Unlike non-degenerate thin-shell wormholes, whose dynamics is governed by the Israel junction conditions, the present model employs the extended equivalence principle to determine the normal motion of the throat. The resulting one-dimensional Kepler problem admits a canonical quantum formulation with a dynamical symmetry described by the non-compact group \(SU(1,1)\).

The algebraic diagonalization yields the discrete spectrum
\begin{equation}
  E_n^2=M^2\left(1-\frac{M^4}{n^2}\right),
\qquad n=1,2,3,\ldots.
\end{equation}

For sub-Planckian masses \(M<m_P\), all discrete levels correspond to real energies, and the system exhibits a Bohr-type spectrum analogous to that of the hydrogen atom. The characteristic size of the quantum states is determined by the gravitational Bohr radius \(b_B=\hbar^2/(GM^3)\).

As the wormhole mass increases beyond the Planck mass, the lowest spectral values \(E_n^2\) successively become negative. However, this does not imply the disappearance of the corresponding quantum states. In the co-moving quantum description, these states remain part of the discrete spectrum. Their apparent disappearance arises only when the quantum states are interpreted from the perspective of a distant observer, whose classical description requires a real asymptotic time and therefore a nonnegative energy square \(E^2\). States with \(E_n^2<0\) cannot be assigned such a classical asymptotic-time evolution and are consequently classified as hidden states.

This distinction is analogous, in a broader sense, to the appearance of classically inaccessible configurations in ordinary quantum mechanics. Just as quantum theory describes wave functions in classically forbidden regions, the present quantum description retains states that cannot be represented within the classical dynamics accessible to a distant observer. The analogy does not imply that the hidden states are literally localized behind the classical event horizon. Rather, their defining property is the absence of a corresponding real asymptotic-time description.

An interesting geometric feature of this classification is that the transition between visible and hidden states, determined by \(E_n^2=0\), corresponds to the characteristic radius \(\bar b_n=3M\), coinciding with the Schwarzschild photon sphere. Thus, the photon sphere emerges as the boundary between the two spectral sectors, whereas the gravitational radius \(\bar b=2M\) represents the minimum radius of a wormhole as seen by a distant observer.

A gradual decrease of the wormhole mass, possibly associated with Hawking evaporation, would reverse the sequence of spectral transitions, allowing previously hidden levels to reappear in the visible sector. Within the quasi-static approximation, this suggests a transition from a semiclassical non-Bohr regime associated with the Einstein–Rosen geometry to a sub-Planckian Bohr-type wormhole. Describing such an evolution requires not merely a gravitational quantum-mechanical picture, but rather taking into account gravitational quantum-field factors, which lie beyond the scope of this model.

Finally, the transition at \(M=m_P\) bears an analogy to the critical coupling \(Z\alpha=1\) in the relativistic Coulomb problem \citep{granovskii_1,granovskii_2}. In both cases, a spectral boundary signals a change in the applicability of the conventional physical interpretation of quantum states. The group-theoretical connection with supercritical vacuum polarization suggests a possible direction for studying quantum processes in the near-horizon region of degenerate wormholes. A consistent treatment of such effects—including Hawking-like radiation and non-spherical fluctuations —may be an interesting and substantive challenge for the present approach.

\bibliographystyle{elsarticle-harv} 
\bibliography{references}



%% For example:

%% \bibitem[Aladro et al.(2015)]{Aladro15} Aladro, R., Martín, S., Riquelme, D., et al. 2015, \aas, 579, A101

%%\end{thebibliography}

\end{document}